\documentclass[aps,twocolumn,showpacs,floatfix]{revtex4}

\usepackage{amsfonts}
\usepackage{amsmath}
\usepackage{amssymb}
\usepackage{graphicx}
\usepackage{color}
\usepackage{verbatim}
\usepackage{lipsum}
\usepackage{dsfont}

\usepackage[export]{adjustbox}
\usepackage{dcolumn}
\usepackage{bm}
\usepackage[mathlines]{lineno}

\usepackage[T1]{fontenc}

\usepackage{hyperref}
\usepackage{makecell}

\begin{document}

\preprint{APS/123-QED}

\title{Surface magnetic structure investigation of a nanolaminated Mn$_2$GaC thin film using a magnetic field microscope based on Nitrogen-Vacancy centers
}

\author{Andris Berzins$^1$}
\email{andris.berzins@lu.lv}
\author{Janis Smits$^{1,2}$}
\author{Andrejs Petruhins$^3$}
\author{Hugo Grube$^1$}

\affiliation{$^1$Laser Centre, University of Latvia, Latvia}
\affiliation{$^2$The University of New Mexico, Albuquerque, United States}
\affiliation{$^3$Materials design group, Thin film physics division, Department of Physics, Chemistry and Biology (IFM), Linkoping University, Sweden}

\date{\today}

\begin{abstract}
This work presents a magnetic field imaging method based on color centers in diamond crystal applied to a thin film of a nanolaminated Mn$_2$GaC MAX phase. Magnetic properties of the surface related structures have been described around the first order transition at 214 K by performing measurements in the temperature range between 200 K and 235 K with the surface features fading out by increasing temperature above the transition temperature. The results presented here demonstrate how Nitrogen-Vacancy center based magnetic microscopy can supplement the traditionally used set of experimental techniques, giving additional information of microscopic scale magnetic field features, and allowing to investigate the temperature dependent magnetic behavior. The additional information acquired in this way is relevant to applications where surface magnetic properties are of essence.
\end{abstract}

\maketitle


\section{Introduction}

The Nitrogen-Vacancy (NV) centers in diamond crystal have proven themselves as a universal platform for magnetic property research of various origins and applications. When used as a magnetic field microscope, that can reveal two dimensional structures of the sample, NV layers can be used for magnetic mapping of superconducting thin films~\cite{waxman_diamond_2014, xu_mapping_2019} and microscopic electromagnetic devices~\cite{pham_magnetic_2011, mizuno_simultaneous_2020,horsley_microwave_2018}, as well as imaging of micrometer and nanometer sized particles~\cite{smits_estimating_2016,fescenko_diamond_2019} and nanoscale magnetic resonance~\cite{ziem_quantitative_2019} and magnetic spins ~\cite{steinert_magnetic_2013} to name a few. Furthermore, these measurements can be done over a wide temperature range ~\cite{waxman_diamond_2014, wang_coherent_2020, plakhotnik_luminescence_2010}. Thus, it is evident that NV based microscopy is well suited for investigation of temperature dependent magnetic effects on microscopic scale.

In this study we demonstrate how NV center based imaging can be a powerful tool that gives significant insight in magnetic properties of the surface of a nanolaminated (atomically layered) MAX phase thin films. \textit{M}$_{n+1}$\textit{A}\textit{X}$_n$ phases (\textit{n} = 1, 2, or 3) are layered hexagonal structures~\cite{barsoum_mn1axn_2000}, where \textit{M} is an early transition metal (e.g. Ti, V or Cr), \textit{A} is an A-group element (e.g. Al, Ga, or Ge), and \textit{X} is either C or N. MAX phase materials are of great interest due to a number of significant properties: metallic-like electrical and thermal conductivity and ceramic-like mechanical properties (high stiffness, damage tolerance and resistance
to corrosion and thermal shock)~\cite{barsoum_mn1axn_2000,magnuson_chemical_2017-1}, as well as superconduction~\cite{hadi_superconducting_2020,barsoum_mn1axn_2000}, large magnetostriction~\cite{novoselova_large_2018} and self-healing properties~\cite{farle_conceptual_2015,farle_demonstrating_2016,sloof_repeated_2016} to name a few.

Magnetism is relatively novel property in the family of MAX phases~\cite{mockute_synthesis_2013,ingason_magnetic_2013}. Although active research is still ongoing in finding new magnetic MAX phases, potential applications for these materials could be found in spintronics~\cite{li_control_2019} and refrigeration~\cite{ingason_magnetic_2016}, to name a few. In search of novel magnetic thin film materials with specific magnetic properties and transition temperatures, it is of great importance to secure a material of as high purity as possible with preferably no amounts of impurities, to avoid any possible interference of magnetic signal from strongly magnetic impurities. 

The typical analysis methods used for determination of thin film properties are  X-ray diffraction (XRD), scanning and transmission electron microscopy (SEM and TEM), and either vibrating sample magnetometry (VSM) or Superconducting quantum interference device (SQUID). XRD is very common as it is easy and cost efficient method to analyse the structural properties of crystalline material allowing for identification of crystalline phases and their content in the sample. SEM and TEM are more sophisticated methods that allow to investigate surface morphology, microstructure and also sample composition. And VSM or SQUID is used to determine the bulk magnetic properties of the sample. There is, however, a perilous gap in the interpretation of the results, and this gap is related to magnetic properties of the surface. As this research will demonstrate, the chemical composition, overall magnetic properties and structural properties can satisfactory, but the magnetic response to external magnetic fields of the surface structural impurities can significantly differ from the expected. For example, even under optimal film growth, A element is known to segregate to the surface~\cite{eklund_epitaxial_2011, ingason_magnetic_2013}, thus presence of A element on the surface can act as a crystallization center for impurities, which could be of such crystal orientation or at such low quantity that would not be detected by the XRD, yet could give rise to magnetic signal of unknown origin. 

Here we demonstrate how NV center based magnetic field imaging (NVMI) can be used as a tool to determine surface magnetic structure with a wide field of view that can be all measured simultaneously while maintaining micrometer resolution. This gives the possibility to adjust the thin film growth parameters to avoid unwanted surface structure elements related to magnetic field distortions.

In this research the investigated material was Mn$_2$GaC, a thin film that has been studied before~\cite{novoselova_large_2018,ingason_magnetic_2013}.
Based on these previous studies, it was decided to monitor the magnetic properties around the  first order phase transition at 214 K, thus a range from 200 K to 235 K was chosen.

\section{Methods used}

\begin{figure}[ht]
  \begin{center}
    \includegraphics[width=0.35\textwidth]{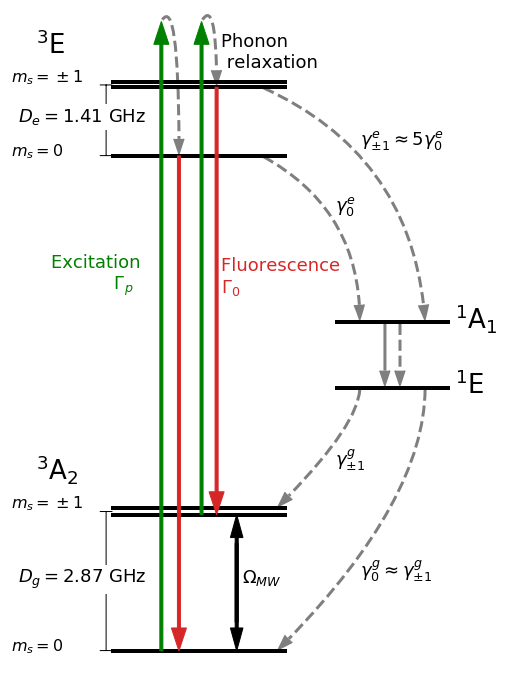}
  \end{center}
  \caption{Level scheme of an NV center in a diamond crystal, $m_S$ is the electron spin projection quantum number, $D_g$ and $D_e$ are the ground state and excited state zero magnetic field splittings, $\Omega_{\text{MW}}$ is the MW Rabi frequency, $\gamma_0^g$ and $\gamma_{\pm 1}^g$ are the relaxation rates from the singlet state $^1$E to the triplet ground state $^3$A$_2$, $\gamma_0^e$ and $\gamma_{\pm 1}^e$ are the relaxation rates from the triplet excited state $^3$E to the singlet state $^1$A$_1$ \cite{auzinsh_hyperfine_2019}.}
  \label{levels}
\end{figure}

To measure the magnetic field distribution we used the optically detected magnetic resonance (ODMR) signals. This method relies on excitation and relaxation cycles represented in Figure \ref{levels}. We used a 532 nm laser light to optically excite the NV centers. After light absorption with rate $\Gamma_p$ and rapid relaxation in the phonon band the excited state magnetic sublevels can either decay back to the ground state with equal rates $\Gamma_0$ or undergo non-radiative transitions to the singlet level $^{1}A_1$. The non-radiative transitions from the excited state electron spin sublevels $m_S = \pm 1$ occur about five times more rapidly than transitions from $m_S = 0$. Then the singlet--singlet transition ${^{1}A}_1 \longrightarrow$ $^{1}E$ takes place with almost all the energy being transferred in a non-radiative way with a small portion being IR radiation (1024 nm). Finally, non-radiative transitions from $^{1}E$ to the ground triplet state have roughly equal transition probabilities to all three electron spin projection components of the $^3A_2$ level. It has to be noted that in the literature data the relaxation rates vary in a rather wide range~\cite{auzinsh_hyperfine_2019, jaskula_superresolution_2017, dumeige_magnetometry_2013},
yet all mentioned publications agree that differences of the non-radiative transition rates in the excited triplet state $^3E$ leads to the situation, where the population of the NV centers in the ground triplet state will be transferred to the magnetic sublevel $m_S = 0$ after several excitation--relaxation cycles. This, in turn, leads to situation where the electron spin angular momentum is strongly polarized \cite{doherty_nitrogen-vacancy_2013}.

The optical polarization is exploited in the following way. By adding microwave (MW) frequency scan in the range where the $m_S = 0$ and $m_S = +1$ or $m_S = -1$ are in resonance, the population eventually is transferred between $m_S = 0$ and $m_S = +1$ or $m_S = -1$ state. As the luminescence from the ground states $m_S = \pm 1$ is lower, this leads to luminescence drop. This gives the possibility to measure the magnetic field as the luminescence signal drops to the minimum value only when the MW frequency is equal to the distance between $m_S = 0$ and $m_S = +1$ or $m_S = -1$. If the external magnetic field is aligned along one of the NV axes the $m_S = \pm 1$ states of the corresponding NV direction split linearly with growing magnetic field by $28.025$~\textit{~\textit{MHz/mT}} ~\cite{doherty_theory_2012}. As the luminescence signal was collected and detected using an optical system, that maintains the two-dimensional information of the luminescence distribution of the NV layer, the magnetic field distribution over the field of view could be reconstructed after the data processing.

\begin{figure}[h]
      \includegraphics[width=0.48\textwidth]{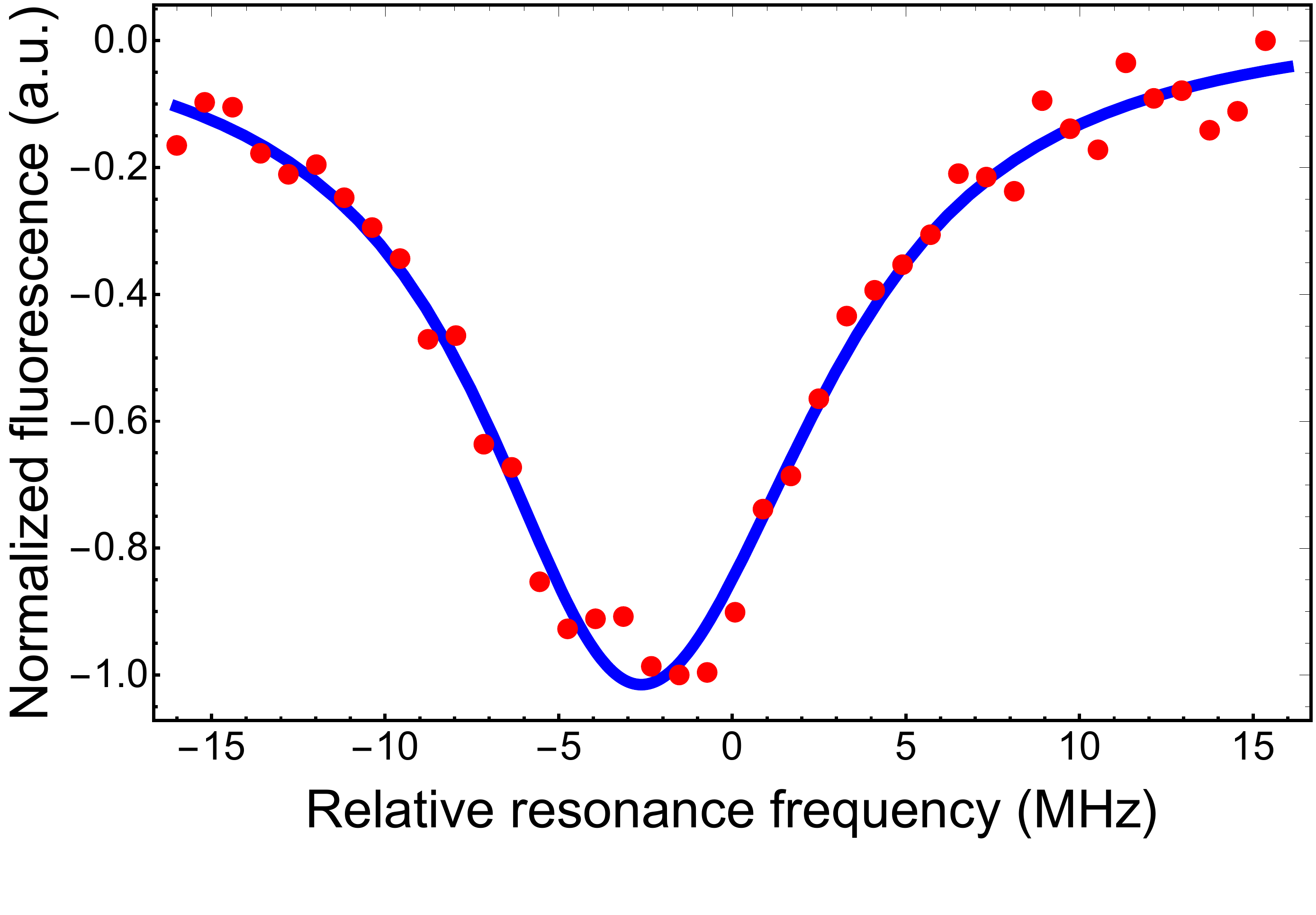}
    \caption{An example of data for one camera pixel: red dots - experimentally measured data points that represent a narrow MW frequency range, that are fitted with a combination of three Lorentzian profiles  - blue line.}
  \label{data}
\end{figure}

The acquired data (full ODMR shape for each pixel with 40 data points) was processed in the following way. The acquired data points were fitted with a combination of three Lorentzian profiles (corresponding to each of the three hyperfine components) for each pixel, where each experimental point represents a narrow MW field value. This was done to account for the asymmetry in the line-profile (see Figure~\ref{data}) due to polarization of the nitrogen nucleus forming the NV center~\cite{doherty_nitrogen-vacancy_2013, neumann_excited-state_2009}. Determining the minimum value from the fitted curve lessens the imperfections as noise and relatively sparse data points at the tip of the resonance shape in data. The obtained fit values for the transition frequencies (frequency at the minimum value of the fit curve) were plotted as a 2D colour map revealing the spatial distribution of the magnetic field (represented in Results and Analysis section  Fig.~\ref{ODMR_maps} and Fig.~\ref{correlation}).

\section{Experimental Setup}

\begin{figure}
      \includegraphics[width=0.48\textwidth]{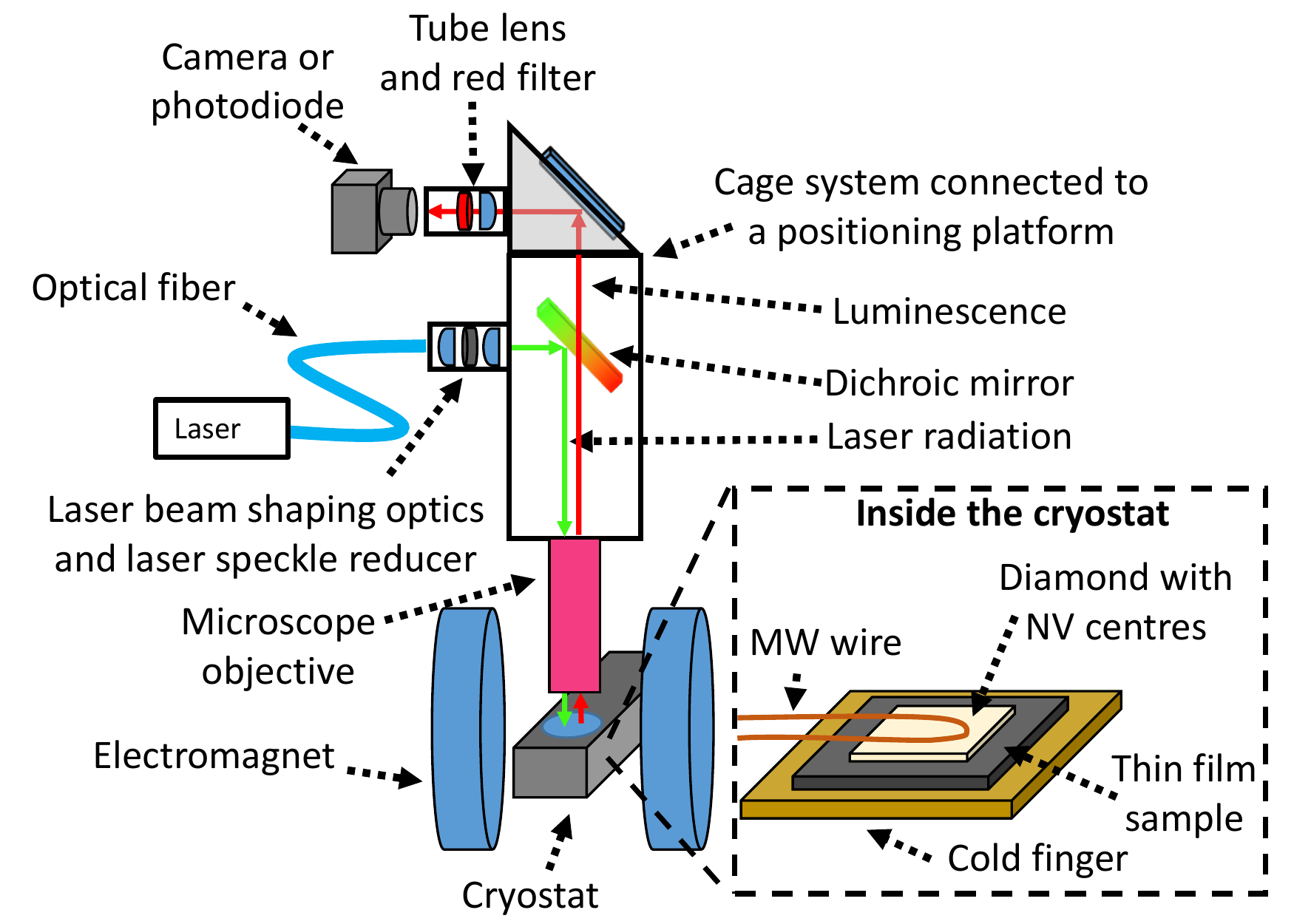}
    \caption{Experimental setup. The laser light was coupled into an optical fiber that led it to the cage system mounted on a positioning platform. The lens system in combination with laser speckle reducer ensures more even distribution of illumination of the NV layer. The dichroic mirror (Thorlabs DMLP567) reflects the green light, which, in turn, is directed to the sample via microscope objective. The luminescence from the sample is collected with the same microscope objective, and, after passing through the dichroic mirror, tube lens and a long-pass filter (Thorlabs FEL0600), it is focused onto a sCMOS matrix of the camera (Andor NEO 5.5) or a photodiode during system adjustments. Inset \textbf{Inside the cryostat} shows the combination of the MW wire, the diamond crystal, the sample and the cyostat cold finger, as used during measurements.}
  \label{setup}
\end{figure}

The experimental setup for NVMI is depicted in Figure~\ref{setup}. We used a Coherent Verdi V-18 laser to excite the NV centers. Laser power that reached the NV layer was around 100 ~\textit{mW}. We used a dynamic transmissive speckle reducer (Optotune LSR-3005) to suppress interference artifacts originating from various sources. A lens system was used to optimise the lighting over the field of view. Due to specifics of the cryostat we used a long working distance infinity corrected microscope objective (Mitutoyo LCD Plan Apo NIR 50X, NA=0.42, F=200). Using an epifluorescent setup the excitation and luminescence detection are done through the same optical path. The red luminescence was separated by a dichroic mirror (Thorlabs DMLP567) in combination with long-pass filter. After that the luminescence was collected either on sCMOS sensor of Andor Neo 5.5 camera, or, during adjustments of system, on the photodiode (Thorlabs PDA36A-EC). The MW field, created by MW generator (SRS SG386) in combination with MW amplifier (Mini Circuits ZVE-3W-83+), was delivered to the NV centers by using a copper wire. The whole optical system was mounted on a positioning platform (Thorlabs NanoMax) that allowed coarse positioning by using stepper motors and during measurements the position was corrected using the piezoelectric actuators.

Measurement process to obtain two dimensional ODMR maps were done by sweeping the MW frequency while simultaneously acquiring a series of camera frames. The MW frequency modulation was controlled by an analog voltage generated by a data acquisition card (NI USB-6001). The sweep was triggered by a pulse generated by the camera itself and a fixed burst of frames were acquired in sync with the modulation waveform (a saw-tooth wave). 40 frames were acquired for each sweep and 5 sweeps were done between each readout. The 5 separate sweeps were then averaged together. This  was done  to  optimize downtime  from  shuttling  data  in  memory.

The field of view of the measurements was a 70x70 $\mu m^2$ square. In all measurements the bias magnetic field was aligned along one of the NV axes (110 orientation), meaning that it is oriented parallel to the diamond surface and thin film sample surfaces. The NVMI experiments were conducted in an electromagnet originally designed for Electron Paramagnetic Resonance measurements, thus it provided highly homogeneous magnetic field.

The temperature of the thin film sample was controlled by using liquid nitrogen cooled cryostat (Oxford Instruments), which provided temperature stability of $\pm~0.2~K$.  The measurements were done in 3.1$\cdot{10^{-6}}$ $\pm$~0.2$\cdot{10^{-6}}~mbar$ vacuum (Edwards T-Station 85H Wet).

The diamond sample used (bought from from Sumitomo Electric) was an HPHT diamond with a (110) surface polish. The crystal dimensions are 2~$mm$ $\times$ 2~$mm$ $\times$ 0.5~$mm$. The diamond crystal was irradiated with $^{4}$He$^+$ ions at three separate energies: 5~$keV$ (1.30$^{12}$ $ions/cm^2$), 22.5~$keV$ (1.60$^{12}$ $ions/cm^2$) and 53~$keV$ (2.60$^{12}$ $ions/cm^2$) and then annealed at high temperature so that the vacancies created by the irradiation could migrate and form NV centers. According to the SRIM simulations~\cite{ziegler_srim_2010} and simple diffusion model this should create a NV layer approximately 250~$nm$ thick and located at one of the surfaces of the diamond crystal.

\section{Preparation and investigation of the thin film}

The detailed description of preparation of the Mn$_2$GaC thin film can be found in another publication~\cite{ingason_magnetic_2013}. 
Surface morphology and composition of the thin film was characterized using a LEO 1550 scanning electron microscope (SEM) equipped with energy dispersive X-ray spectroscopy (EDX).

To perform the NVMI procedure the MgO substrate with the deposited Mn$_2$GaC thin film, the thin film sample was glued to the nonmagnetic cold finger of the cryostat by using a layer of Crystalbond 509 glue. After that the diamond crystal edges were fastened to the sample by using a carbon tape in a way that the NV layer faces the thin film. After that the MW wire was put on top of the diamond crystal (aggregation can be seen in the inset of the Figure~\ref{setup}). To secure the distance between the thin film, diamond and MW wire, we used a simple aluminium clamp, that gently pushed everything together. This aggregate was designed in such a way that it still allowed optical access from the above (through the window of the cryostat).

\section{Results and analysis}

The temperature range, that was chosen for analysis in present study was set around the first order magnetic transition of nanolaminated Mn$_2$GaC~\cite{novoselova_large_2018} at 214~$K$. In this transition, material transforms from antiferromagnetic (AFM) at higher temperatures to non-collinear AFM at lower temperatures, where the non-collinear state yields a resulting ferromagnetic component. This phase transition happens gradually over a temperature range starting from 180 K and ending at 240 K. And this defined the temperature range of interest, and, as will be explained later, helped to identify the chemical composition of the structures, that created the local magnetic field distortions.

The measurements in present study were done at constant magnetic field of about 45 $mT$, and at several temperatures: 200~$K$, 210~$K$, 213.5~$K$, 217~$K$, 225~$K$ and 235~$K$. To separate the magnetic effects from other measurement artefacts, we did the measurements on both NV ground state transitions $m_S = 0 \longrightarrow$ $m_S = \pm 1$. It is known that only magnetic effects shifts the ODMR resonance frequency of $m_S = +1$ and $m_S = - 1$ in opposite directions (shifting $m_S = +1$ to positive direction, but $m_S = - 1$ to negative direction), while temperature effects shift both frequencies in positive direction. In this case the strain induced ODMR frequency shifts are not of concern, as the thin film is not deposited on the diamond surface.

\begin{figure*}
      \begin{center}
      \includegraphics[width=1\textwidth]{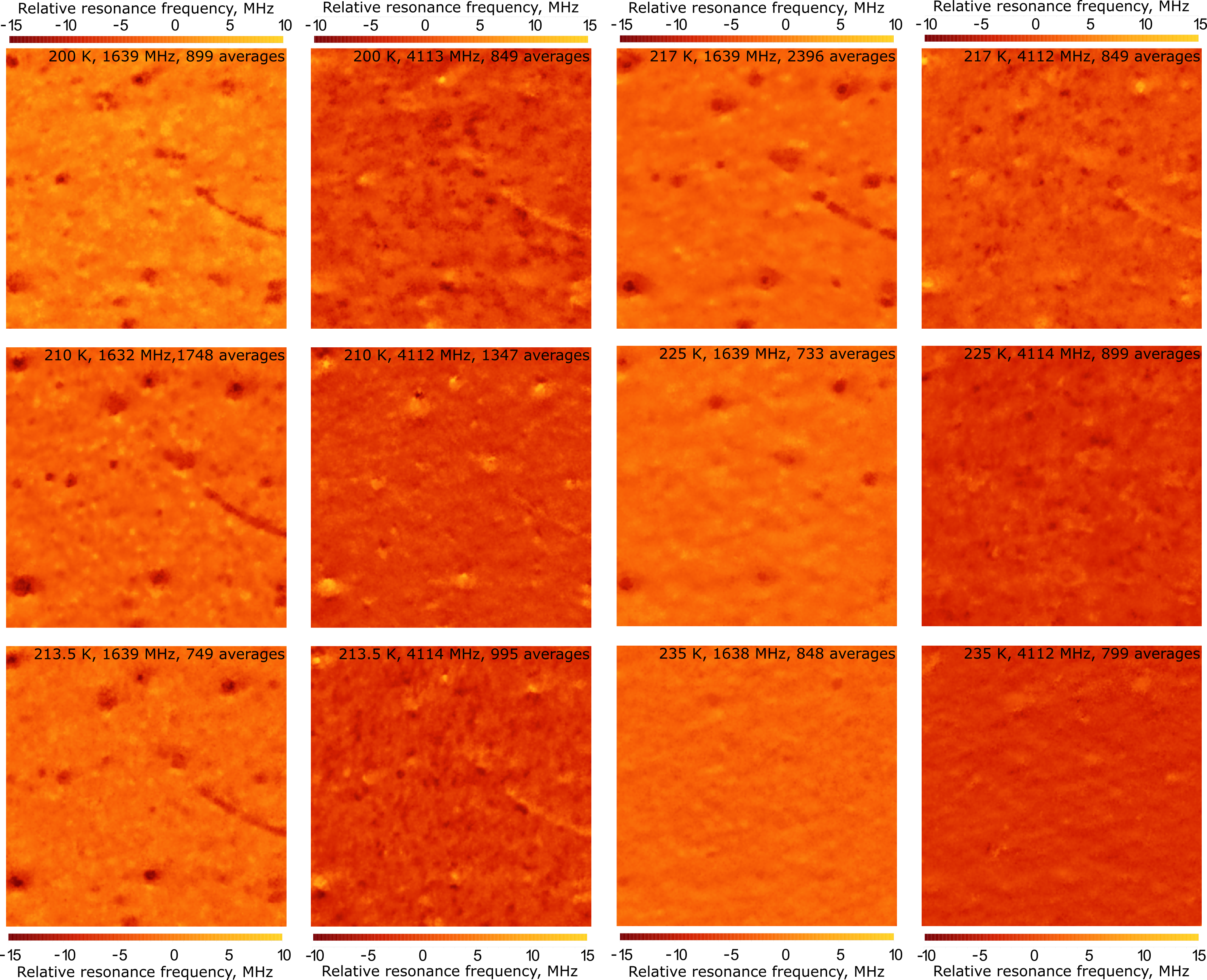}
    \end{center}
    \caption{ODMR maps (magnetic images) measured on $m_S = 0$ to $m_S = -1$ transition (the first and the third column) and $m_S = 0$ to $m_S = +1$ transition (the second and the fourth column) represented in pairs. In both measurement sets it is visible that local magnetic features are pronounced at 217~$K$ and below, while these features start to vanish at 225~$K$, and are almost completely gone at 235~$K$. This temperature dependent magnetic properties corresponds to the expected of Mn$_2$GaC MAX phase. The color scale represents the relative resonance frequency shift with the central resonance frequency of the MW scan in each case is depicted as zero for better readability. The external magnetic field around 45~$mT$ in the direction parallel to the magnetic field maps represented here. The information in the right top corner of each map gives the temperature the measurement was done at, central frequency of the scan and the average count for the measurement.}
  \label{ODMR_maps}
\end{figure*}

The magnetic field maps acquired are shown in Figure~\ref{ODMR_maps}. In these magnetic field maps the main body of the field of view remains relatively uniform regardless of the temperature at which the measurements are conducted, however, there are notable features that do change with the temperature. These peculiar features appear as rapid shifts in the local magnetic field, and they behave in a ferromagnetic manner - containing magnetic signal that has components both in the direction of the applied magnetic field and opposite to it. In addition, these ferromagnetic features have distinct temperature dependent properties; while the features remain unchanged at temperatures below 217~$K$, above these temperatures these features start to fade out, and at 235~$K$ only hardly noticeable traces remain.

The magnetic field maps in Figure~\ref{ODMR_maps} are structured in pairs, the first and the third column represents transition from $m_S = 0$ to $m_S = - 1$, and the second and the fourth column represents transition from $m_S = 0$ to $m_S = +1$. By looking at the magnetic field maps it is clear, that various sized features create opposite shifts of the MW resonance frequency - in the case of transition from $m_S = 0$ to $m_S = - 1$ the features shift the resonance frequency to the negative side (red) of the central frequency, while in the case of transition from $m_S = 0$ to $m_S = +1$ the features shift the resonance frequency to the positive side (yellow) of the central frequency. This means that their origin is magnetic.

\begin{figure*}[ht!]
      \begin{center}
      \includegraphics[width=1\textwidth]{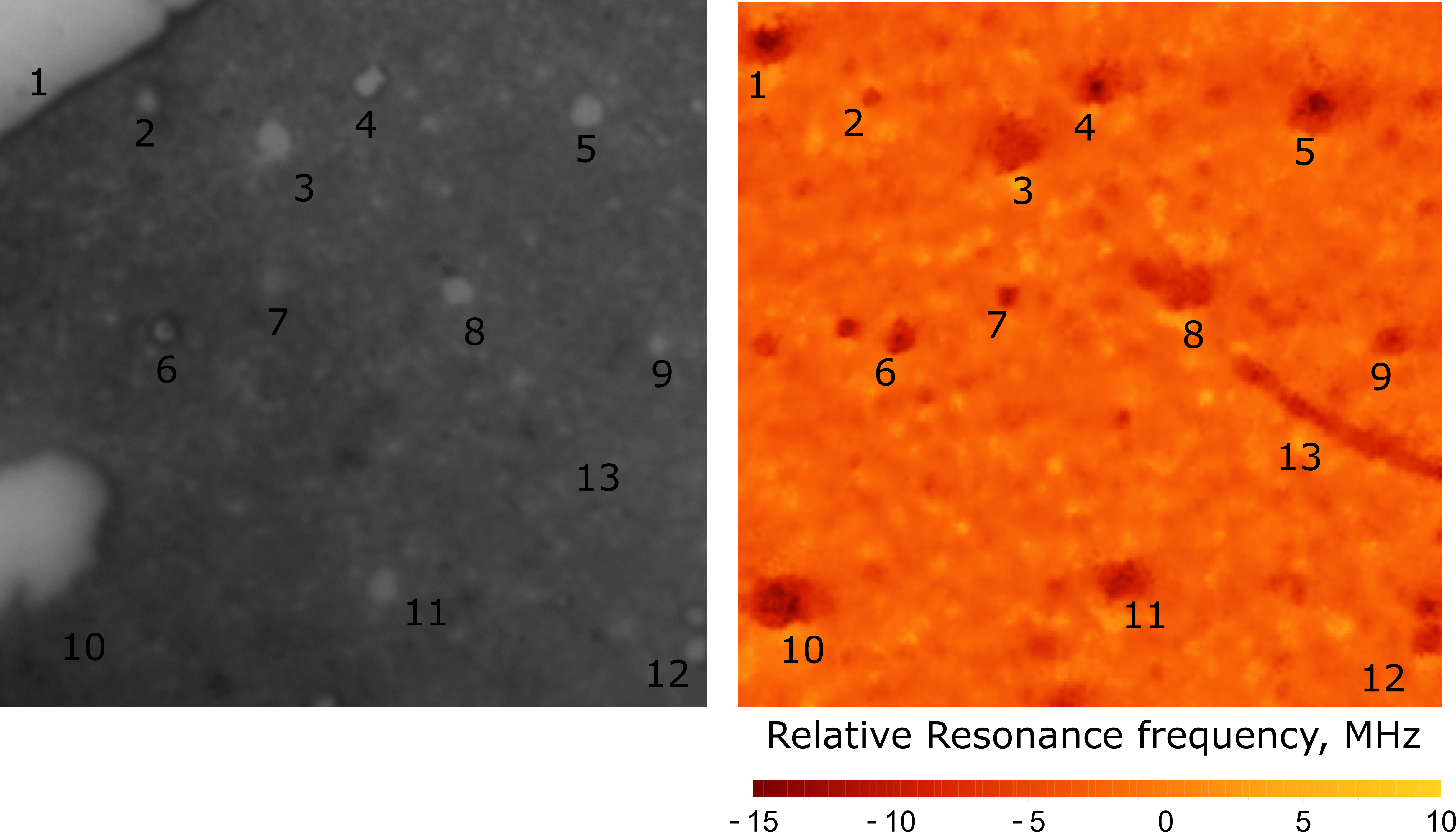}
    \end{center}
    \caption{The field of view of images is 70 $\times$ 70 $\mu$m. {\textbf{Left panel}} - Optical image of the surface of the thin film illuminated by a white light. The lighter regions represent surface impurities. The large light grey regions are fragments of 100 $nm$ Al layer, used to demonstrate the possibility to detect obscured defects as features 1 and 10. Feature 13 has close to no signature in the optical image, but there is a group of small surface impurities that most probably are responsible for the magnetic signal. The \textbf{right panel} represents the magnetic field map obtained at 210~$K$. The color scale represents the relative resonance frequency shift with the central resonance frequency of the scan range 4122~\textit{MHz} depicted as zero for better readability}
  \label{correlation}
\end{figure*}

To better understand the origin of the ODMR signals, we correlate the magnetic images with the optical image of the same region - Figure~\ref{correlation}. It has to be noted, that the optical image was taken through the relatively thick diamond crystal, with which the magnetic field images were made. Thus, the image lacks the sharpness that could be achieved by taking picture outside the region covered with diamond. The clearly pronounced features are numbered 1-13 in Figure~\ref{correlation}. The features 2-9, 11 and 12 are clearly related to micrometer sized features that appear brighter than the rest of the film in optical image. Features 1 and 10 represent the possibilities of the NV based imaging to detect obscured magnetic features, as they are located beneath a 100 $nm$ thick aluminium mirror fragments, that are sputtered on the surface of the diamond. The line-like feature marked as 13 consists of a combination of smaller sized features, that collectively form seemingly uniform line-like structure. The surface of this thin film is covered with numerous different sized surface features, and we assigned a number to only small fraction of the most pronounced features. From the viewpoint of magnetic properties, the thin film acts as a homogeneously magnetized plate, and the features on the surface are alone responsible for the local magnetic field direction changes, which in turn resembles the response of a magnetic sphere.

After detailed analysis of the magnetic surface features by the SEM and EDX, we attribute the above described features to arbitrary oriented micrometer sized crystal grains. In EDX it could be confirmed that these tiny crystals contain Mn, Ga and C, same as the bulk of the thin film sample. The origin of these structures is deduced to be related to formation of gallium droplets on the surface of the thin film, attributed to A element segregation to the surface, previously observed for other MAX phases~\cite{eklund_epitaxial_2011, ingason_magnetic_2013}. These droplets seem to stimulate crystal growth on them. It has to be noted that the chemical composition of these surface crystals is also confirmed by their temperature dependent magnetic properties: the diminishing magnetism at temperatures nearing 240~$K$. Moreover, from by the chemical composition of these crystals, there are no known compounds in the compositional region of Mn-Ga-C system have magnetic phase transitions at these temperatures.

It should be added that the magnetic surface features, if compared at temperatures of 217~$K$ and below at some images seem to be slightly more focused, and in some slightly defocused. This happens due to several reasons, first, it is relatively hard to determine the central frequency of the whole field of view when the measurements are done, thus there are some deviations in what is depicted as relative zero (determined by central frequency of MW scan during measurements), and what is the real relative zero on the ODMR scale (the central ODMR frequency, when all the field of view is averaged together). This leads to a slight variation in the assignment of concrete color to a concrete offset from center frequency, that also leads to some cases being red in a wider region, for example. The second thing in this perspective is the measurement process itself, the position on the thin film surface is selected and during measurements maintained manually. As it is done by visually pinpointing some points of reference to a reference grid, there are possible small variations (around few micrometers large) in the position of the features. This applies to a single measurement, and also to cases where several measurements are averaged together (cases with average count above 1000 averages), leading to few micrometer smearing out of the magnetic field features. 

\section{Discussion}

The obtained experimental results clearly demonstrate the possibilities of NVMI on the microscale. We have shown clearly identifiable detection of features on the surface of Mn$_2$GaC thin film around the first order magnetic phase transition at 214 K. Here we would like to add that due to mode of operation the magnetic response of the main thin film might seem to be very minor, but this is only because we focused on the surface features. In the presentation of the data it led us of choosing the relative ODMR frequency map frequency centers, color scales and color saturation in such a way that background information is almost lost, but the surface features are highlighted.

The detected magnetic surface features convincingly match the magnetic behavior of the Mn$_2$GaC, but due to the geometrical placement and size, respond to the external magnetic field in a different manner. These surface features would not change the magnetic behavior of the thin film as a whole in this case for a bulk measurement but strong local magnetic features on the surface might be of concern for applications where surface magnetic properties are of essence.

In this research have presented the different advantages of NVMI for possible research applications - opportunity to compare optical images with magnetic images with microscopic resolution, or, as demonstrated, opportunity to measure magnetic signals of visually obscured features. Furthermore, although the spatial resolution and magnetic field sensitivity of this method cannot compete with different scanning tip magnetometers, the possibility to measure the hole field of view at the same time with sub-micrometer resolution gives this method great potential for high throughput applications.

\section{Acknowledgements}

A. Berzins acknowledges support from PostDoc Latvia project 1.1.1.2/VIAA/1/16/024 "Two-way research of thin films and NV centres in diamond crystal" and LLC "MikroTik" donation, administered by the UoL foundation, for opportunity to significantly improve experimental setup.

\bibliography{references}

\end{document}